\documentclass[12pt]{article}

\usepackage{times}

\usepackage[dvips]{graphicx}
\usepackage[usenames,dvipsnames]{color}

\usepackage{amssymb}

\usepackage[T1]{fontenc}
\usepackage{calc}
\usepackage{tabularx}
\usepackage{longtable}
\usepackage{setspace}
\usepackage{color}
\usepackage{bar}
\usepackage{url}

\begin{document}

\title{Dynamic Linking of Smart Digital Objects Based on User Navigation Patterns}
\author { Aravind Elango, Johan Bollen and Michael L. Nelson\\
	Computer Science Department\\
	Old Dominion University\\
	Norfolk Virginia USA - 23529\\
	\{aelango,jbollen,mln\}@cs.odu.edu	
	}

\maketitle

\abstract{	
	We discuss a methodology to dynamically generate links among digital objects by means
	of an unsupervised learning mechanism which analyzes user link traversal patterns.
	We performed an experiment with a test bed of 150 complex data objects, referred to as buckets.
	Each bucket manages its own content, provides methods to interact with users and individually
	maintains a set of links to other buckets.  We demonstrate that buckets were capable of dynamically adjusting
	their links to other buckets according to user link selections, thereby generating a meaningful
	network of bucket relations. Our results indicate such adaptive networks of linked buckets approximate the collective
	link preferences of a community of users.}

\section{Introduction}
Current research in the area of recommender systems has focused on analyzing static representations
of user preferences, e.g. list of purchased items, to generate personalized recommendations \cite{up:reco}.
However, user preferences are not static and shift as users assume different roles and interests. Furthermore,
digital library applications often do not concern purchasable items but complex information objects, 
and user interests must be inferred from less explicit statements of interest. One such mechanism of 
inferring user interests is to analyze the links previously traversed by the user. We use buckets, smart
digital objects, which individually manage a dynamic list of links to other buckets, to generate run-time, adaptive recommendations.

\subsection{Smart Objects: Buckets}

Buckets are smart objects for the aggregation of data \cite{bucket:nelson2001}.
Buckets contain mechanisms to aggregate, manage, protect and preserve the data they contain.
A bucket could be thought of as an intelligent, active folder which, among 
other functionalities, also contains interface methods to display its contents.\\

Buckets are not simply passive, folder-like repositories: they have an internal structure.
A bucket may contain 0 or more elements, each of which can contain elements in their own rights.
An element may be a resource such as a PDF file, a data set or simply a set of other elements.
An element may be a ``pointer'' to any arbitrary network object, e.g. another bucket,
in the form of a URL.  By having an element ``point'' to other buckets,
buckets can logically contain other buckets.\\

Buckets have no predefined size limitations, either in terms
of storage capacity, or in terms of number of elements. Authors can model whatever application
domain they desire using the basic structure of elements. Bucket methods can be activated by user
HTTP requests.\\

As an example of how methods in an bucket are invoked, consider the bucket identified by the URL:\\\\
\url{http://www.cs.odu.edu/~mln/naca-tn-2509/}\\

When no bucket method is specified, the ``display'' method is assumed. Therefore the mentioned URL is
equivalent to: \\\\
\url{http://www.cs.odu.edu/~mln/naca-tn-2509/?method=display}\\

The above mentioned URLs will induce the bucket to return an overview of the elements it contains.
These elements themselves can again be URLs containing requests for bucket methods.
A specific bucket method allows a bucket to redirect a request for its content to another object,
which could be another bucket. For example, the request:\\

\url{http://www.cs.odu.edu/~mln/naca-tn-2509/?method=display&re\ direct=http://naca.larc.nasa.gov/reports/1951/naca-tn-2509/}\\

would request the odu.edu bucket to redirect to the nasa.gov bucket.
All requests for external resources are first routed through the bucket that contains the link.
The full bucket API is discussed in \cite{bucket:nelson2000}.\\

 The motivation for buckets came from previous experience in the design, implementation and maintenance 
 of NASA scientific and technical information Digital Libraries (DLs), including the NASA Technical Report Server (NTRS)
 \cite{ntrs:nelson1995}. Buckets are well suited for distributed applications because they can aggregate 
 heterogeneous content and remain functional in low-fidelity environments. Since they are self-contained, 
 independent and mobile, they should be resilient to changing server environments. In addition, buckets 
 can be adapted to a variety of data types and data formats.

\subsection{Adaptive user interfaces}

Hebb's law of learning \cite{organi:hebb1949}, an essential component of many unsupervised methods in machine learning, 
is the basis of our efforts to generate meaningful and dynamic sets of inter-bucket links. 
We have used a descriptive methodology \cite{role:smith1993} to bucket linking,
where the user interface changes based on the past 
actions of users and not on predictions of users' future actions. An advantage of such adaptive user interfaces
is that they can dynamically take into account user information needs by continuously updating their structure
and presentation.

\subsection{Hebb's Law of Learning}

Hebb's law postulates that the connection between two neurons in the human brain becomes stronger
when the neurons are persistently activated in quick succession to one another. As such the brain continuously
adapts the connections between neurons based on previous experiences. Although Hebb's law represents a coarse
and incomplete picture of neural plasticity, it has found countless applications in machine learning. Hebb's law
is specifically applicable to situations in which no set of correct or erroneous responses can be defined in advance,
and the system needs to gradually acquire information which is only implicitly present in a given data set.\\

For this reason, Hebbian learning has been successfully used in adaptive hypertext networks \cite{system:bollen1998}
which learn to reroute hyperlinks according to usage patterns. In analogy to such systems, we use a variation of Hebbian
learning for dynamic inter-bucket linking.

\section{Implementing Hebb's laws in buckets}
\begin{figure}
       \begin{center}
       \includegraphics[scale=0.7]{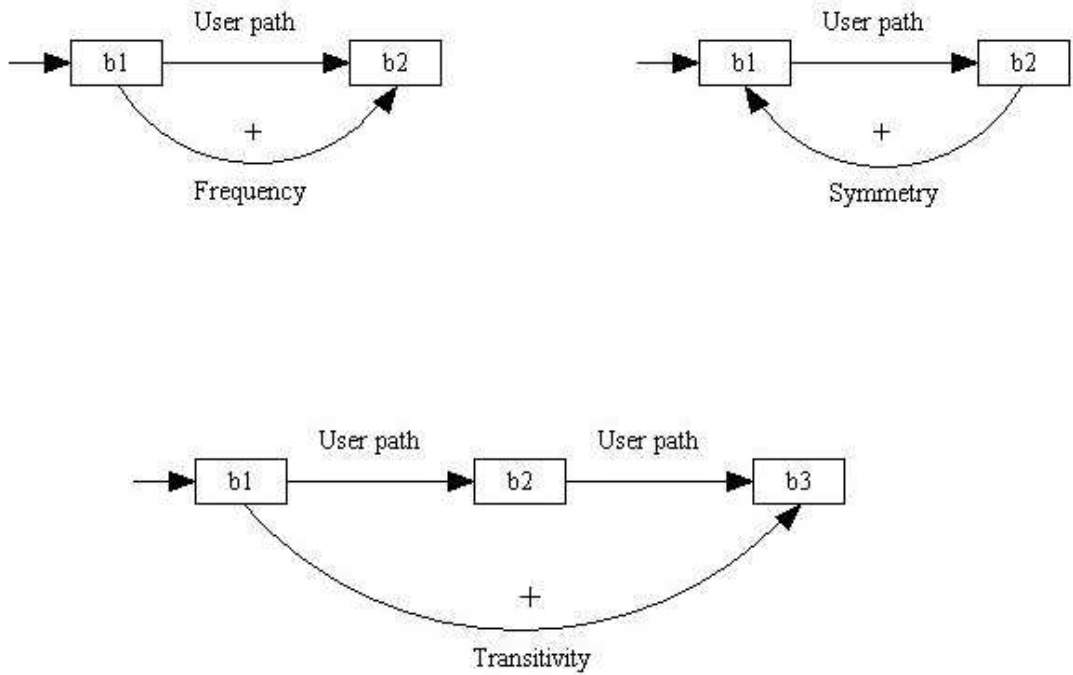}
       \end{center}
       \caption{\label{hebbian} Implementing Hebbian learning in buckets}
 \end{figure}

Fig. \ref{hebbian} gives an overview of how Hebbian learning can be interpreted for inter- 
bucket linking. Let us imagine the user traversing 3 buckets namely b1, b2 and b3. b1 is linked to b2 and
b2 is linked to b3. It is assumed that there are no other links among these 3 buckets initially. When a 
user traverses from b1 to b2, the link ($b1 \rightarrow b2$) is strengthened by a frequency reinforcement. When b2 is linked to
from b1, we conjure that b2 is related to b1 and strengthen the link ($b2 \rightarrow b1$). If the link ($b2 \rightarrow b1$)
is absent, it is created. When the user traverses from b2 to b3, the weight of the link ($b2 \rightarrow b3$) is 
increased by a frequency reinforcement and the weight of the link ($b3 \rightarrow b2$) is incremented by a symmetry reinforcement.
Since the user finally reached b3 from b1 with b2 as an intermediary, we assume that b1 has some degree of 
relation to b3 and hence we increase the strength of the link ($b1 \rightarrow b3$) by a transitivity reinforcement. If 
the link ($b1 \rightarrow b3$) is absent, it is created.\\

The approach taken to implement the above procedures was
suggested in \cite{adapti:bollen2002}. When a bucket b1 is expected to link to bucket b2, b1 is called with a redirect argument and this argument 
gives the URL of the bucket it is linking to. We also pass a referer argument which essentially
overrides the HTTP referer argument. The values passed by the referer argument would be instrumental in
implementing Hebb's laws as explained below.\\

When the link from b1 to b2 is traversed, b1 is called with the URL: \\
\url{http://b1?method=display&referer=b1&redirect=http://b2?method\ =display%26referer=http://b1} \\\\
b1 knows itself as a referer by 
seeing the referer argument and also concludes by seeing the redirect argument that it is redirecting 
to b2.  Thus the link to b2 in b1 is incremented by a given frequency reinforcement. b2 sees that the referer 
is b1 and increments the weight of its link ($b2 \rightarrow b1$) by a given symmetry weight. When the user
next traverses to b3 from b2, the following link is dynamically generated:  \\

\url{http://b2?method=display&referer=b2&redirect=http://b1?method\ =display%26redirect=http://b3?method=display%26referer=http://b2.} \\\\
b2 sees itself as the referer and finds that b3 is the 
final destination based on the last redirect argument. b2 increases the weight of its link to b3 by a given frequency 
reinforcement. After incrementing the link weight, b2 redirects to b1. b1 sees that there 
is no referer argument and so, increases the link weight of ($b1 \rightarrow b3$) by the transitivity reinforcement.
Finally when b3 is called, it finds the referer argument to be b2 and increments the weight of 
the link ($b3 \rightarrow b2$) by the symmetry reinforcement.\\

Reinforcement values are based on our experiences with previous systems: the frequency, symmetry and transitivity reinforcements are respectively
defined as 1.0, 0.5 and 0.3. The frequency weight is the highest since the user directly traverses this link and we have positive confirmation
that this link was deemed relevant.

\section{Experimental Test Bed}

\begin{figure}
       \begin{center}
       \includegraphics[scale=0.5]{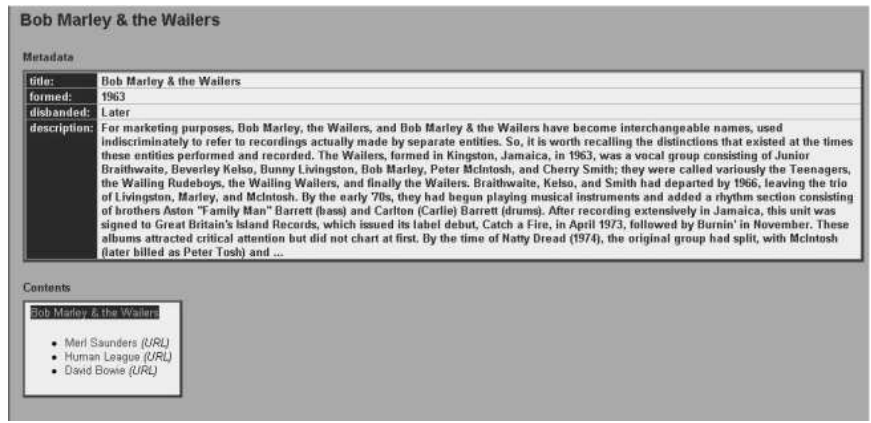}
       \end{center}
       \caption{\label{display} An example of bucket display}
 \end{figure}

One hundred and fifty buckets were used for this experiment. Each bucket represented a popular 
music artist, containing a short biography of the band and a dynamic list of related links to other buckets in the network.
The list of 150 music bands was composed from the top 50 bands of all times as chosen by experts from Spin Magazine 
\cite{url:spin} and two each of their similar bands as suggested in www.allmusic.com.
Each bucket was initially randomly linked to 3 other buckets with a weight of 0.5 to provide an initial
unbiased navigation structure.  Fig. \ref{display} shows the display in one such bucket.
The bucket displays metadata related to the band and a set of links to other artists/bands. As users traverse the 
system, new links are created and the weights of pre-existing links are increased based on the users link 
selection. The set of links are sorted based on their weight so that a heavily weighted link is shown higher 
up in the list of links than a less weighted link.\\

\begin{figure}
       \begin{center}
       \includegraphics[scale=0.7]{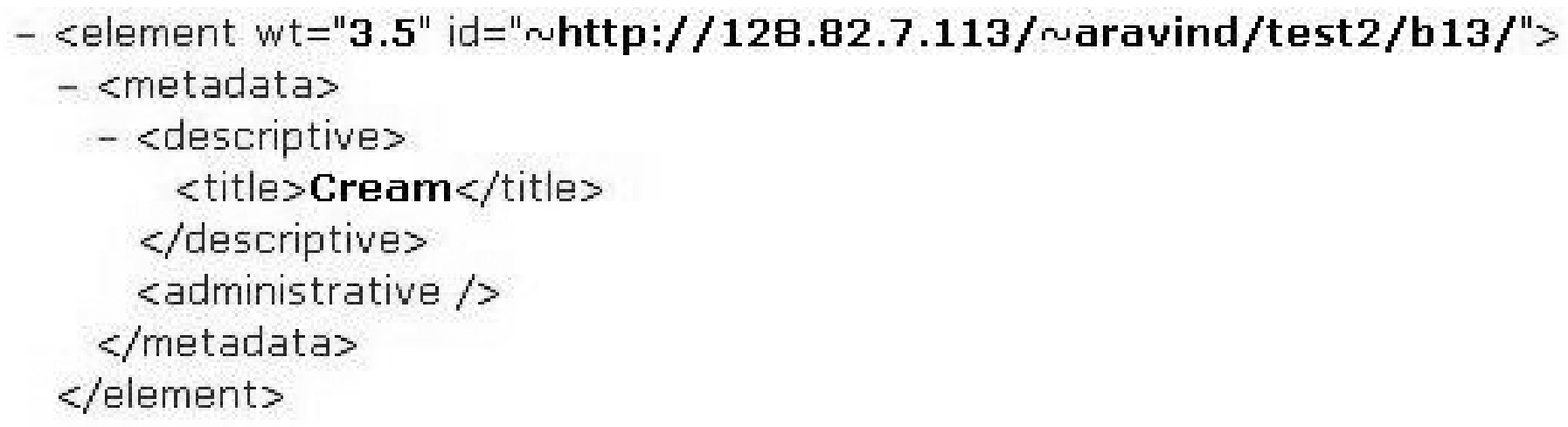}
       \end{center}
       \caption{\label{xml} XML representation of an element and the associated weight}
 \end{figure}

Every bucket has an XML file, which contains all information about the elements
in the bucket and their metadata. The weight of each individual link is stored as an attribute of the
element (URL it is pointing to) in the XML file, as shown in Fig. \ref{xml}. New elements can be added to 
the XML file using the addElement method.\\

An invitation to traverse the network was sent to 15 people in June 2003. The total weight associated 
with all the links excluding the initial random links was 1719 units at the end of the experiment.
Taking into account the reinforcements assigned for frequency, symmetry and transitivity we estimate the system to have had approximately
1041 direct traversals. 

 \begin{table}
 \begin{center}
 \begin{tabular}{|c||c|}

 \hline
\multicolumn{2}{l}{\textit{Example Bucket: 'The Clash'}} \\ \hline
\textbf{Links before traversal}		&	\textbf{Links after traversal} \\ \hline & \\
 \textcolor{red}{The Beatles} 	& Smashing Pumpkins 	\\ 
 \textcolor{red}{Glyn Jones } 	& \textcolor{red}{Beck }  	\\ 
\textcolor{red}{Beck } 	& Fishbone \\
			& Nick Lowe \\
			&  \textcolor{red}{The Beatles}\\
			& The Smiths \\ 
			& Replacements \\
			&  \textcolor{red}{Glyn Jones }\\
			& N.W.A \\
			& Squeeze \\ & \\
 \hline
 \end{tabular}
 \caption[short title here]{\label{the_clash} An example of the dynamic links generated for `The Clash'.}
 \end{center}
 \end{table}

 \begin{table}
 \begin{center}
 \begin{tabular}{|c||c|}

 \hline
\multicolumn{2}{l}{\textit{Example Bucket: 'The Smiths'}} \\ \hline
\textbf{Links before traversal}		&	\textbf{Links after traversal} \\ \hline & \\
 \textcolor{red}{Elvis Costello} 	& \textcolor{red}{Replacements }\\ 
 \textcolor{red}{Tool} 	& \textcolor{red}{Elvis Costello } 	\\ 
\textcolor{red}{Replacements} 	&	Pretty Things 	\\ 
			& Fishbone \\
			& Nick Lowe \\
			& The Beatles\\
			& Fishbone \\ 
			& The Clash \\
			&  Johnny Thunders\\
			& Kiss \\
			& \textcolor{red}{Tool } \\ & \\
 \hline
 \end{tabular}
 \caption[short title here]{\label{the_smiths} An example of the dynamic links generated for `The Smiths'.}
 \end{center}
 \end{table}

\section{Results}

Our aim is to prove that when users start surfing the collection from a portal node or bucket, a meaningful network 
develops in which the content of highly centric nodes is similar or related to the content of the portal bucket.

\subsection{Link Structure}
The bucket representing `Public Enemy' was the entry point to the network. This setup is similar to a portal
through which users access web services. (e.g. www.yahoo.com). 
The portal bucket and every heavily traversed bucket starts reflecting the users preference on what other 
buckets should be linked to the current bucket. Table \ref{the_clash} shows the links associated 
with `The Clash' bucket before and after traversal. As users navigate the bucket, new links are dynamically 
created and the bands which users presume are more related to `The Clash' bubble up to the top.\\

Another similar example is shown in Table \ref{the_smiths}. It is evident that users do not associate `Tool' ( 
an initial random addition) with `The Smiths' and hence it has dropped down in the list of links as compared
to `Replacements' and `Elvis Costello' which are also random initial links but have maintained their positions
at the top of the list, and do match the nature of The Smiths as a music band.

\subsection{Bucket Authority}

A highly influential node within a network can be expected to have a relatively high number of outgoing and incoming
connections to and from other nodes in the network, a characteristic refered to as degree centrality.
An investigation into the degree centrality of nodes in a network will reveal the network's most important nodes. Applied
to the generated network of buckets,  degree centrality can therefore be used to partially validate network structure.
Since our buckets concern music bands, degree centrality may relate to the relative importance or influence of
music bands according to the community of users that generated the network.\\

We define the degree centrality of a node as the number of links that originate from or terminate 
in that particular node. The weighted degree centrality of a node is computed as the sum of all the weights of the links 
that originate from or terminate in that particular node. \\

Degree centrality $dc_i$ is defined as Eq.\ref{degree_centrality} where $l_{ij}$ is 1 if there exists a link
from bucket $i$ to $j$, zero otherwise. Weighted degree centrality $wc_i$ is defined as Eq.
\ref{wt_degree_centrality} where $w_{ij}$ is the weight of the link linking bucket $i$ to bucket $j$. 
$w_{ij}$ is 0 if bucket $k$ is not linked to bucket $j$.

\vspace*{1cm}

\begin{equation}
{dc_i = \sum_{j=1}^n l_{ij} +  \sum_{j=1}^n l_{ji}}
\label{degree_centrality}
\end{equation}

\begin{equation}
{wc_i = \sum_{j=1}^n w_{ij} +  \sum_{j=1}^n w_{ji}}
\label{wt_degree_centrality}
\end{equation}

\vspace*{1cm}

\begin{figure}
\begin{center}
\input{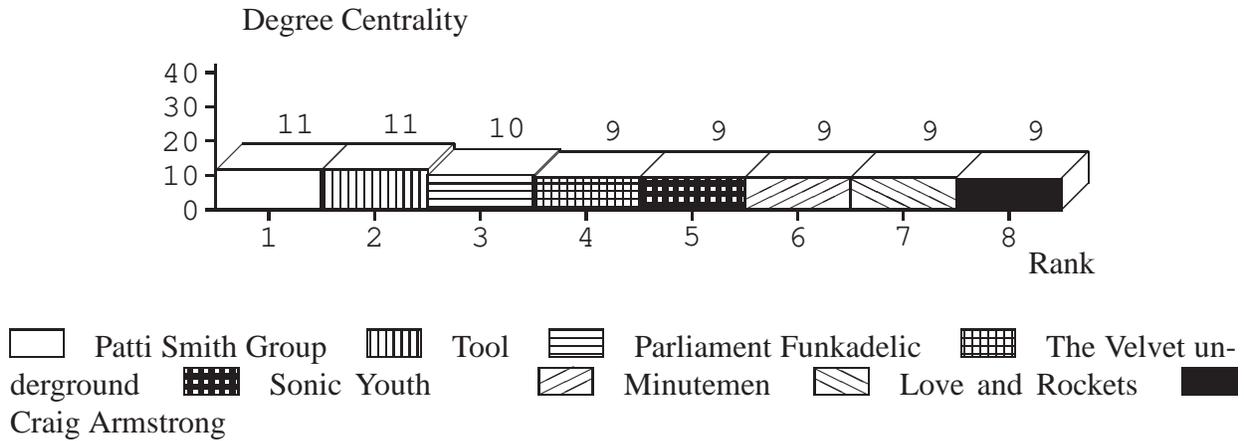}
\caption{\label{initial_deg_bar} Top eight degree centrality rankings based on initial random linking.}
\end{figure}

\begin{figure}
\begin{center}
\input{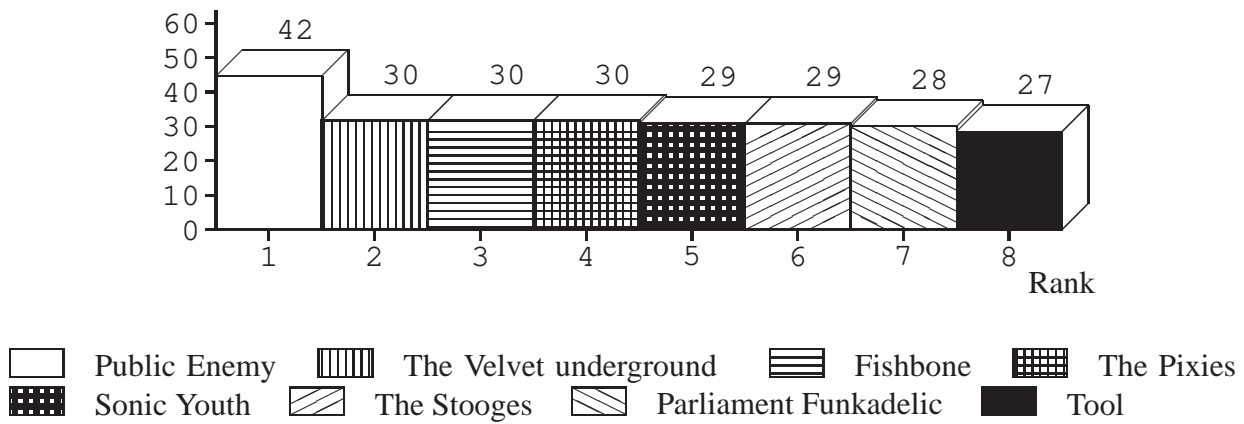}
\caption{\label{final_deg_bar} Top eight degree centrality rankings after users surfed the system.}
\end{figure}

Fig. \ref{initial_deg_bar} shows the ranking of the top 8 buckets based on degree centrality rankings when
the network was intially setup. The rankings in this case are purely random and no user traversals had taken 
place. In this case, the degree and weighted degree centrality rankings are the same.
Fig. \ref{final_deg_bar} shows the top 8 rankings based on degree centrality after approximately 1041 direct
traversals by 15 users. `Public Enemy' is seen to be the most popular band according to degree and weighted
degree centrality measures. This was expected since `Public Enemy' was the access point to the network for 
all users. We also find influential bands such as the ``The Velvet Underground'', ``The Stooges'' and ``L.L. Cool J.''

\begin{figure}
\begin{center}
\input{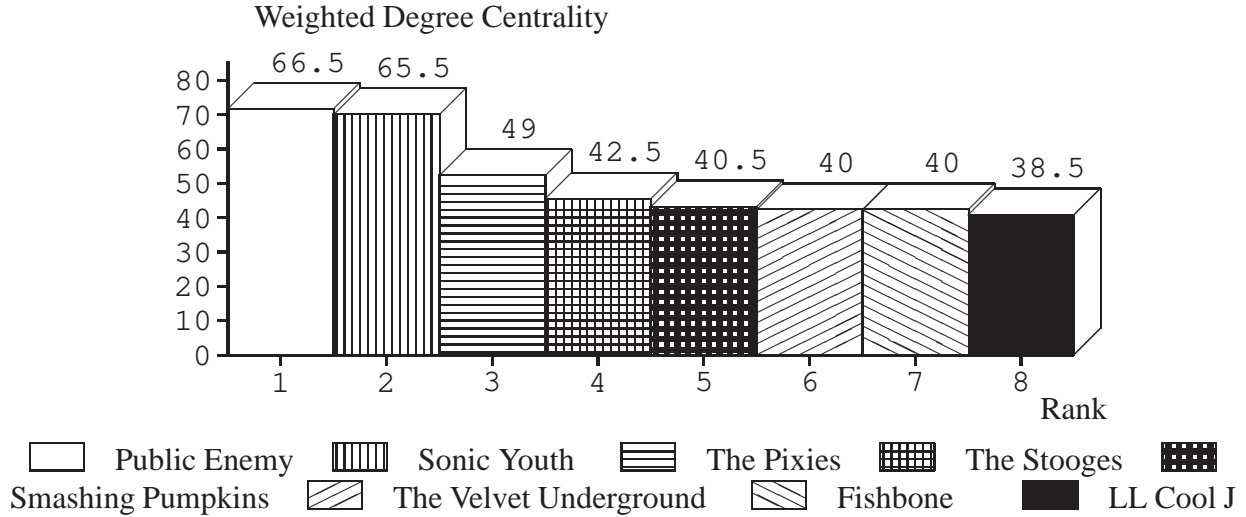}
\caption{\label{final_wt_bar} Top eight weighted degree centrality rankings after users surfed the system.}
\end{figure}

\subsection{Hierarchical Ranking}

\begin{figure}
       \begin{center}
       \includegraphics[scale=0.5]{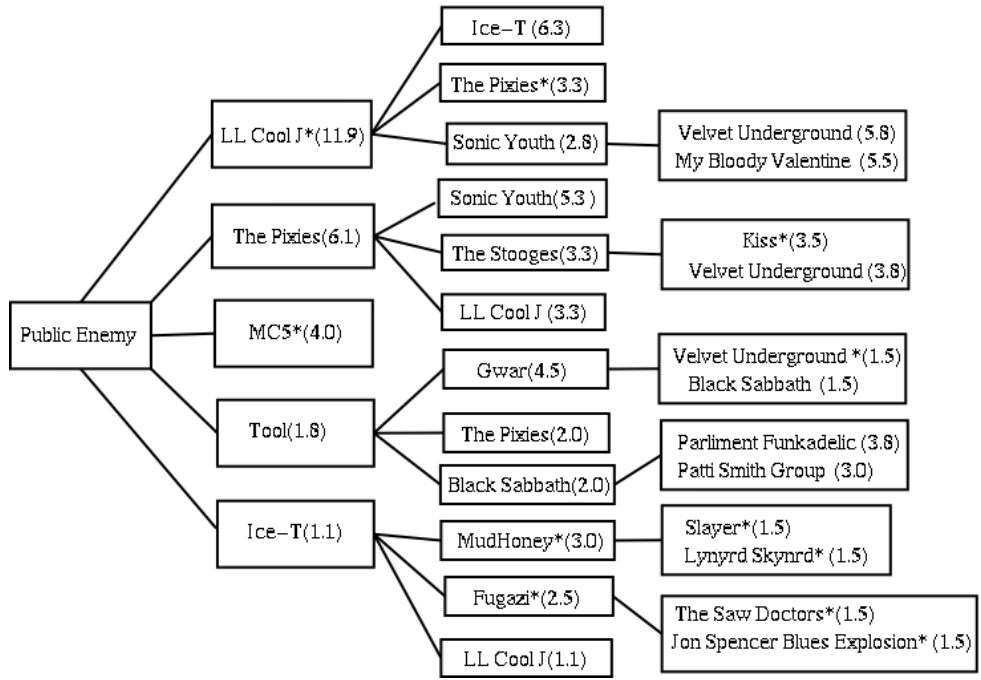}
       \end{center}
       \caption{\label{hierarchical_graph} Hierarchical ranking of bands related to `Public Enemy'.}
 \end{figure}

Since all users entered the network starting at the ``Public Enemy'' buckets, it makes sense to investigate
this buckets connections to other buckets as a means to validate network structure.
Fig. \ref{hierarchical_graph} gives the hierarchy of the most popular bands starting from ``Public Enemy'' 
including secondary, tertiary and reinforced but initially random links. The weight of the links 
connecting every two bands is noted within parentheses next to the band lower in the hierarchy.
Each band marked with * indicates that it is an initial random link which has been reinforced.\\ 

The 150 bands were graded by two music experts on a scale from 10 to 0, with 10 signifying close relation between the 
band and ``Public Enemy'' and 0 signifying no relationship between the bands. The rankings were later 
normalized to a scale of 1. We compute the relationship weights between every band in Fig. \ref{hierarchical_graph} and 
`Public Enemy' as the sum of the product of all normalized intermediary link weights in order to compare
network weights to the expert opinion.\\

We can formalize this procedure as follows. Assume two buckets $b_i$ and $b_j$ are connected in the network shown in Fig. \ref{hierarchical_graph}
via a path $p$ of length $n$, so that the ordered set $p=\left(b_1, b_2, \cdots, b_k\right)$ represents
the buckets on the path that connects $b_i$ and $b_j$. Multiple paths can be identified between any two
buckets, therefore we have a set of $k$ paths $P = \{p_1, p_2, \cdots, p_k\}$.\\

Eq. \ref{link_wt} is used to compute the weight of relationship of any bucket $b_i$ and the bucket $b_j$ in
the generated hierarchical tree, given that $W\left(b_h \in p_g, b_{h-1} \in p_g\right)$ represents the weight between
the bucket $b_h$ and its predecessor $b_{h-1}$ in path $p_g$.

\begin{equation}
W(b_i, b_j) = \sum_{g=1}^k \prod_{h=2}^n W( b_h \in p_g, b_{h-1} \in p_g)
\label{link_wt}
\end{equation}

We examined the indirect link weights of any buckets and the ``Public Enemy'' bucket, so $b_i$ is 
assumed to be the ``Public Enemy'' bucket in all cases.\\

\begin{figure}
\begin{center}
\input{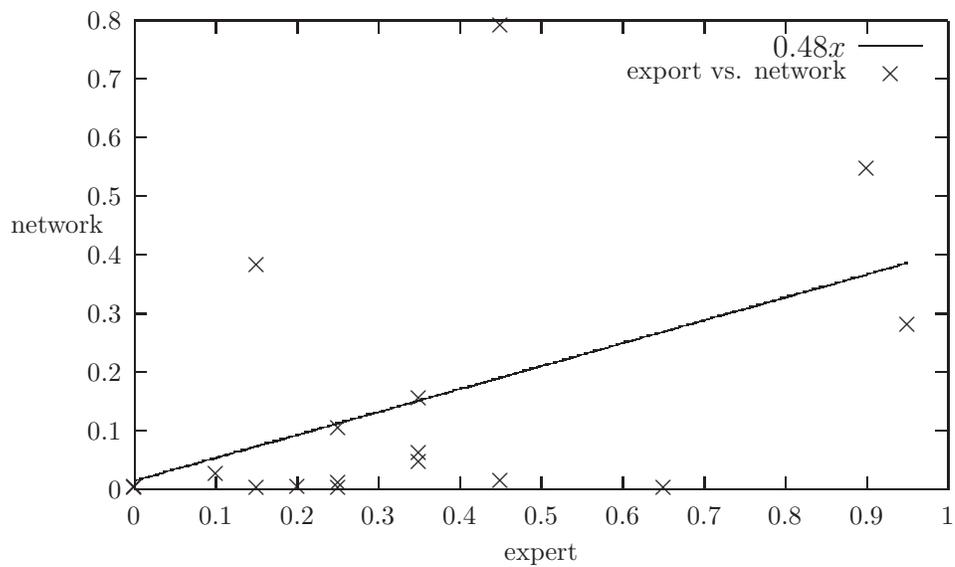}
\end{center}
\caption{\label{scatter_jb_net} Comparison of expert and network ranking of band relationships to `Public Enemy'.}
\end{figure}

Fig. \ref{scatter_jb_net} shows the scatterplot of the expert and network relationship values to `Public Enemy'.
The correlation coefficient between network and expert evaluations of bucket relations to ``Public Enemy''
was found to be 0.48 indicating that relationships in the graph correspond to at least two expert judgments. 
The expert opinions need further validation and with more usage the network could be expected to better reflect
 user tastes.

\section{Future Research}

While the importance of the nodes has been gauged based on degree and weighted degree 
centrality measures, it would be interesting to perform an analysis based on principal components 
and other clustering techniques. \\ 

Another system feature could be decrementing the weight of rarely used links. 
This would help filter out spurious links created by the initial random linking.
The basis for decrementing the link weight needs further study. Options include the
time for which a link has not been accessed and the frequency of access of other links. \\

Finally, Hebbian learning could be implemented on a portion of a bucket instead of the entire bucket. 
This would allow a bucket to have section(s) of content that are fixed and section(s) of content
that can adapt by Hebbian learning. Imagine bucket1 containing 2 high level
elements/sections (HLE1 and HLE2) each with a number of leaf elements. When the user links from 
bucket1 to bucket2 via a link provided in HLE1, bucket2 would be aware of not only the bucket
it was linked from (bucket1) but also the section (HLE1) within that bucket.

\section{Conclusion}

We have implemented a system for the automated linking of information using a collection
of smart objects, labeled buckets, using a set of simple learning rules which change link weights 
based on user retrieval patterns. The bucket networks gradually change structure as users
retrieve one bucket after another via a list of recommended buckets. \\

It is evident from the results that although a collection of buckets are initially randomly linked, 
with adequate user traversal they form a meaningful linkage with resembles the users idea of which 
buckets should be related to which other buckets. The most centric nodes in the network happen to be 
either influential or very popular music bands related to ``Public Enemy''.  These bands have high 
degree and weighted degree centralities. \\

It was found in the course of analysis, that the rankings based degree centrality was more susceptible 
to change due to drastic use even by a single user. However, weighted degree centrality offers a more 
graded and stable approach.\\

The random collection initially presented to the users would not return the ideal results needed to satisfy
the users information need. The possibility of the system returning the ideal answer set for a user's
information need increases with usage of the system. The usage needed to create a well suited network
depends on the number of buckets in the network and also on the diversity of the users' information need. 
When the users information need is of limited scope (e.g. all users interested  in rock music)
a meaningful network can be expected to form fairly quickly.

\bibliographystyle{plain}
\bibliography{proj}

\end{document}